# GRAVITATIONAL WAVES, DARK ENERGY AND INFLATION

WEI-TOU NI

*Center for Gravitation and Cosmology, Department of Physics,*
*National Tsing Hua University, Hsinchu, Taiwan, 30013 Republic of China*
*weitou@gmail.com*

In this paper we first present a complete classification of gravitational waves according to their frequencies: (i) Ultra high frequency band (above 1 THz); (ii) Very high frequency band (100 kHz – 1 THz); (iii) High frequency band (10 Hz – 100 kHz); (iv) Middle frequency band ( 0.1 Hz – 10 Hz); (v) Low frequency band (100 nHz – 0.1 Hz); (vi) Very low frequency band (300 pHz – 100 nHz); (vii) Ultra low frequency band (10 fHz – 300 pHz); (viii) Hubble (extremely low) frequency band (1 aHz – 10 fHz); (ix) Infra-Hubble frequency band (below 1 aHz). After briefly discussing the method of detection for different frequency bands, we review the concept and status of space gravitational-wave missions --- LISA, ASTROD, ASTROD-GW, Super-ASTROD, DECIGO and Big Bang Observer. We then address to the determination of dark energy equation, and probing the inflationary physics using space gravitational wave detectors.

## 1. Introduction

In 1905, the special theory of relativity was formulated by Poincaré[1] and Einstein.[2] In the same paper,[1] Poincaré attempted to develop a relativistic theory of gravity and mentioned gravitational-wave propagating with the speed of light based on Lorentz invariance. After many attempts to develop a relativistic theory of gravity by various authors including Einstein himself,[3,4] Einstein proposed the theory of general relativity in 1915 as the relativistic theory of gravitation and explained the Mercury perihelion advance anomaly.[5] Soon after his proposal of general relativity, Einstein predicted the existence of gravitational waves and estimated its strength.[6]

The existence of gravitational waves is the direct consequence of general relativity and unavoidable consequences of all relativistic gravity theories with finite velocity of propagation. Maxwell's electromagnetic theory predicted electromagnetic waves. Einstein's general relativity and relativistic gravity theories predicts the existence of gravitational waves. Gravitational waves propagate in space-time forming ripples of space-time geometry.

The role of gravitational wave in gravity physics is like the role of electromagnetic wave in electromagnetic physics. The importance of gravitational-wave detection is twofold: (i) as probes to explore fundamental physics and cosmology, especially black hole physics and early cosmology, and (ii) as a tool in astronomy and astrophysics to study compact objects and to count them, complement to electromagnetic astronomy and cosmic ray (including neutrino) astronomy.

The existence of gravitational radiation is demonstrated by binary pulsar orbit evolution.[7] In general relativity, a binary star system should emit energy in the form of gravitational waves. The loss of energy results in the shrinkage of the orbit and shortening of orbital period. In the thirty years of observations of the relativistic binary pulsar B1913+16, the cumulative shift of peri-astron time is over 35 s. The calculated orbital decay rate in general relativity using parameters determined from pulsar timing observations agreed with the observed decay rates. From this and a relative acceleration correction due to solar system and pulsar system motion, Weisberg and Taylor[7] concluded that the measured orbital decay is consistent at the $(1.3 \pm 2.1) \times 10^{-3}$ level with the general relativistic prediction for the emission of gravitational radiation.

The usual way of detection of gravitational wave (GW) is by measuring the strain $\Delta l/l$ induced by it. Hence gravitational wave detectors are usually amplitude sensors, not energy sensors. The detection of GWs can be resolved into characteristic frequencies. The conventional classification of gravitational-wave frequency bands as given by Thorne[8] in 1995 is into (i) High-frequency band (1-10 kHz); (ii) Low-frequency band (100 µHz - 1 Hz); (iii) Very-low-frequency band (1 nHz -100 nHz); (iv) Extremely-low-frequency band (1 aHz - 1 fHz). This classification is mainly according to frequency ranges of corresponding types of detectors/detection methods: (i) Ground GW detectors; (ii) Space GW detectors; (iii) Pulsar timing methods; (iv) Cosmic Microwave Background (CMB) methods. In 1997, we follow Ref. [8] and extend the band ranges to give the following classification:[9,10]

(i) High-frequency band (1-10 kHz);
(ii) Low-frequency band (100 nHz - 1 Hz);
(iii) Very-low-frequency band (300 pHz-100 nHz);
(iv) Extremely-low-frequency band (1 aHz - 10 fHz).

Subsequently, we added the Very-high-frequency band and the Middle-frequency band for there were enhanced interests and activities in these bands. Recently we added the missing band (10 fHz – 300 pHz) and the 2 bands beyond to give a complete frequency classification of GWs.[11-14] In Sec. 2, we present this classification and review the corresponding types of GW detectors or the observation methods briefly.

In Sec. 3, we review space GW detectors in more details. In Sec. 4, we discuss the use of GW luminosity distance to test dark energy models. In Sec. 5, we address to the issue of inflationary GWs and the use of space GW detectors to detect them. In Sec. 6, we present an outlook.

## 2. Classification of Gravitational Waves

Similar to frequency classification of electromagnetic waves to radio wave, millimeter wave, infrared, optical, ultraviolet, X-ray and γ-ray etc., we can have the following complete Frequency Classification of Gravitational Waves:

(i) Ultra high frequency band (above 1 THz): Detection methods include Terahertz resonators, optical resonators, and ingenious methods to be invented.
(ii) Very high frequency band (100 kHz – 1 THz): Microwave resonator/wave guide detectors, optical interferometers and Gaussian beam detectors are sensitive to this band.

- (iii) High frequency band (10 Hz – 100 kHz): Low-temperature resonators and laser-interferometric ground detectors are most sensitive to this band.
- (iv) Middle frequency band (0.1 Hz – 10 Hz): Space interferometric detectors of short armlength (1000-100000 km).
- (v) Low frequency band (100 nHz – 0.1 Hz): Laser-interferometer space detectors are most sensitive to this band.
- (vi) Very low frequency band (300 pHz – 100 nHz): Pulsar timing observations are most sensitive to this band.
- (vii) Ultra low frequency band (10 fHz – 300 pHz): Astrometry of quasar proper motions are most sensitive to this band.
- (viii) Extremely low (Hubble) frequency band (1 aHz – 10 fHz): Cosmic microwave background experiments are most sensitive to this band.
- (ix) Beyond Hubble frequency band (below 1 aHz): Inflationary cosmological models give strengths of GWs in this band. They may be verified indirectly through the verifications of inflationary cosmological models.

There are two kinds of GW detectors – (i) the resonant type: GW induces resonances in detectors (metallic bars, metallic spheres, resonant cavities, …) to enhance sensitivities; (ii) detectors measuring distance change using microwave/laser/X-ray… between/among suspended/floating test bodies. Two crucial issues in GW detection are (i) to lower disturbance effects: suspension isolation and drag-free to decrease the effects of surrounding disturbances; (ii) to increase measurement sensitivity: capacitive sensing, microwave sensing, SQUID transducing, optical sensing, X-ray sensing…

In the very high frequency band (100 kHz – 1 THz), there are cavity/waveguide detector,[15] 0.75 m arm length laser interferometer[16] and Gaussian beam detector.[17]

Most of the current activities of gravitational wave detection are in the high frequency band. In this band, the cryogenic resonant bar detectors have already reached a strain spectral sensitivity of $10^{-21}$ $Hz^{-1/2}$ in the kHz region. NAUTILUS put an upper limit on periodic sources ranging from $3.4 \times 10^{-23}$ to $1.3 \times 10^{-22}$ depending on frequency in their all-sky search.[18] The AURIGA-EXPLORER-NAUTILUS-Virgo Collaboration applied a methodology to the search for coincident burst excitations over a 24 h long joint data set.[19] The MiniGRAIL[20] and Schenberg[21] cryogenic spherical gravitational wave detectors for omnidirectional detection are in active commissioning stages.

Major detection efforts in the high frequency band detection are in the long arm laser interferometers. The TAMA 300 m arm length interferometer,[22] the GEO 600 m interferometer,[23] and the kilometer size laser-interferometric gravitational-wave detectors --- LIGO[24] (two 4 km arm length, one 2 km arm length) and VIRGO[25] all achieved their original sensitivity goals basically. Around the frequency 100 Hz, the LIGO and Virgo sensitivities are all in the level of $10^{-23}$ $(Hz)^{-1/2}$. Various limits on the GW strains for different sources become significant. For example, analysis of data from the third and fourth science runs of the LIGO and GEO600 gravitational wave detectors set strain upper limits on the gravitational wave emission from 78 radio pulsars with the lowest as $2.6 \times 10^{-25}$ for PSRJ1603-7202, and limits on the equatorial ellipticities with the lowest less than $10^{-6}$ for PSR J2124-3358; the strain upper limit for the Crab pulsar is about 2.2 times greater than the fiducial spin-down limit.[26] Analysis of the data from a LIGO two-year science run constrain the energy density of the stochastic gravitational-wave background normalized by the critical energy density of the Universe, in the frequency band around 100 Hz, to be $6.9 \times 10^{-6}$ at 95% confidence.[27] This search for the stochastic background improves on the indirect limit from the Big Bang nucleosynthesis at 100 Hz.

Space interferometers (LISA,[28] ASTROD,[29,30] ASTROD-GW,[12,14] Super-

ASTROD,[31] DECIGO,[32] and Big Bang Observer[33,34]) for gravitational-wave detection hold the most promise with signal-to-noise ratio. LISA[28] (Laser Interferometer Space Antenna) is aimed at detection of low-frequency ($10^{-4}$ to 1 Hz) gravitational waves with a strain sensitivity of $4 \times 10^{-21}/(Hz)^{1/2}$ at 1 mHz. There are abundant sources for LISA, ASTROD and ASTROD-GW: galactic binaries (neutron stars, white dwarfs, etc.). Extra-galactic targets include supermassive black hole binaries, supermassive black hole formation, and cosmic background gravitational waves. A date of LISA launch is hoped for 2020. More discussions will be presented in the next section.

For the very-low-frequency band, the ultra-low-frequency band and the extremely-low-frequency band, it is more convenient to express the sensitivity in terms of energy density per logarithmic frequency interval divided by the cosmic closure density $\rho_c$ for a cosmic background of gravitational waves, i.e., $\Omega_g(f)$ (= $(f/\rho_c) d\rho_g(f)/df$).

The upper limits from pulsar timing observations on a gravitational wave background are about $\Omega_g \leq 10^{-7}$ in the frequency range 4-40 nHz,[35] and $\Omega_g \leq 4 \times 10^{-9}$ at $6 \times 10^{-8}$ Hz.[36] Several pulsar timing arrays (PTAs) will improve on the sensitivity for gravitational wave detection.[37] More pulsar observations with extended periods of time will improve the limits by 3 orders of magnitude in the lifetime of present ground and space GW detector projects.

Gravitational waves with periods longer than the time span of observations produce a simple pattern of apparent proper motions over the sky.[38] Therefore, precise measurement of proper motion of quasars would be a method to detect ultra-low frequency (10 fHz – 300 pHz) gravitational waves. Gwinn *et al.*[39] used this method to constrain the gravitational waves with frequency in this band to less than 0.2 closure density. Long baseline optical interferometer with sub-micro-arcsecond and nano-arcsecond (nas) astrometry is technologically feasible.[40] With this kind of interferometer implemented, precision astrometry of quasar proper motions can be improved by 4 order of magnitude and reach nas $yr^{-1}$. In terms of energy, the precision of determining/constraining $\Omega_g(f)$ could reach $10^{-9}$ or better.

Cosmic microwave background experiments are most sensitive to Extremely low (Hubble) frequency band (1 aHz – 10 fHz). The COBE microwave-background quadrupole anisotropy measurement[41,42] gives a limit $\Omega_g$ (1 aHz) ~ $10^{-9}$ on the extremely-low-frequency gravitational-wave background.[43,44] Ground and balloon experiments probe smaller-angle anisotropies and, hence, higher-frequency background. WMAP[45-47] improves on the COBE constraints; the constraint on $\Omega_g$ for the higher frequency end of this band is close to $10^{-15}$. Planck Surveyor[48] space mission is launched last year and can probe anisotropies with $l$ up to 2000 and with higher sensitivity.

In spite of tremendous efforts in high frequency band and very high frequency band experiments, gravitational wave has not been directly detected yet. This is due to the weakness in the strength of gravitational waves.

For a binary of masses $M_1$ and $M_2$ with Schwarzschild radius $R_1$ and $R_2$, the strain h of its emitted gravitational radiation is of the order of

$$h \approx R_1 R_2 / Dd , \qquad (1)$$

where d is the distance between $M_1$ and $M_2$, D the distance to the observer. For neutron star or black hole, d can be of the order of Schwarzschild radius, and estimation can be simplified:

$$h \leq R/D. \qquad (2)$$

For black hole of solar masses, R = 3 km, and D = $10^8$ l.y., h ≤ 3 ×$10^{-21}$; for inspiral of neutral star binaries, the GW strain generated is smaller. The present generation of km-size arm length interferometers reach the sensitivity of detecting binary neutron star inspirals up to the Virgo cluster distance. From the statistics of astrophysical binary neutron star distribution, the chance of detection is about 0.05 events per year. However with a tenfold increase of strain sensitivity, the reach in distance increases by tenfold and the reach in astrophysical volume increases by one thousand, and the chance of detection is about 50 events per year. This is the goal of both Advanced LIGO[49] and Advanced Virgo[50] under construction. When they are completed around 2015, we do expect direct detection of gravitational waves. 3 km LCGT[51] is working on its final approval. The 100 m CLIO paves the road for LCGT cryogenic technology. The public launch of the Australian International Gravitational Observatory/LIGO South[52] has been held recently on February 22, 2010 in Perth, Australia. We may see a global network of second generation km-size interferometers for gravitational wave detection in the later part of this decade.

**3. Space gravitational wave detectors**

The solar-system gravitational field is determined by three factors: the dynamic distribution of matter in the solar system; the dynamic distribution of matter outside the solar system (galactic, cosmological, etc.) and gravitational waves propagating through the solar system. Different relativistic theories of gravity make different predictions of the solar-system gravitational field. Hence, precise measurements of the solar-system gravitational field test these relativistic theories, in addition to enabling gravitational wave observations, determination of the matter distribution in the solar-system and determination of the observable (testable) influence of our galaxy and cosmos. To measure the solar gravitational field, we measure/monitor distance between different nature or artificial celestial bodies. In the solar system, the equation of motion of a celestial body or a spacecraft is given by the astrodynamical equation

$$\mathbf{a} = \mathbf{a}_N + \mathbf{a}_{1PN} + \mathbf{a}_{2PN} + \mathbf{a}_{Gal\text{-}Cosm} + \mathbf{a}_{GW} + \mathbf{a}_{non\text{-}grav} \qquad (3)$$

where $\mathbf{a}$ is the acceleration of the celestial body or spacecraft, $\mathbf{a}_N$ its acceleration due to Newtonian gravity, $\mathbf{a}_{1PN}$ its acceleration due to first post-Newtonian effects, $\mathbf{a}_{2PN}$ its acceleration due to second post-Newtonian effects, $\mathbf{a}_{Gal\text{-}Cosm}$ its acceleration due to Galactic and cosmological gravity, $\mathbf{a}_{GW}$ its acceleration due to GW, and $\mathbf{a}_{nongrav}$ its acceleration from all non-gravitational origins. Distances between spacecraft depend critically on solar-system gravity (including gravity induced by solar oscillations), underlying gravitational theory and incoming gravitational waves. A precise measurement of these distances as a function of time will enable the cause of a variation to be determined. Certain orbit configurations are good for testing relativistic gravity; certain configurations are good for measuring solar parameter; certain are good for detecting gravitational waves. All these are integral part of mission designs.

In 1981, Faller and Bender first studied possible gravitational wave mission concepts in space using laser interferometry.[54,55] In their seminal work, two basic ingredients were addressed – drag-free navigation for reduction of perturbing forces on the spacecraft and laser interferometry for sensitivity of measurement. The present LISA[28] spacecraft orbit formation was reached in 1985 for LAGOS (Laser Antenna for Gravitational-radiation Observation in Space).[56] It is natural for people working in lunar laser ranging and measuring free fall acceleration using interferometry to propose such an experiment. In fact, test mass free fall inside a falling shroud in vacuum in the interferometric measurement of the earth gravitational acceleration can be considered as a passive drag-free navigation device. The discrepancy in the absolute gravimeter comparison at BIPM is partially resolved using correction to interferometric measurements of absolute gravity arising from the finite speed of light.[57] In the spacecraft tracking, finite velocity of light has always been incorporated. Both test mass of LISA and test mass of interferometric gravimeter (lunar tide and solar tide also affect the local gravitational acceleration) can be regarded as freely falling objects in the solar system and tracked using astrodynamical equation (3).

LISA is an ESA–NASA mission for observing low-frequency gravitational waves in the frequency range from $10^{-5}$ Hz to 1 Hz. LISA has nearly equilateral triangular spacecraft formation of arm length $5 \times 10^6$ km in orbit 20° behind earth. It measures time-varying strains in spacetime by interferometrically monitoring variations in the three $5 \times 10^6$ km baselines. The three spacecraft have drag-free control and house two test masses and interferometry equipment. The mission concept has been under study since 1993 and has been in Phase A at both ESA and NASA since 2004. Time delay interferometry is implemented to suppress laser noises. For technological demonstration, LISA Pathfinder is under development and will be launched in 2011. The science goals include gravitational wave observations of (i) supermassive black holes with $10^5$-$10^7$ $M_\odot$ ($M_\odot$: solar mass) masses; (ii) intermediate-mass black holes with $10^2$-$10^5$ $M_\odot$ masses; (iii) extreme-mass-ratio black hole inspirals, e.g., 10 $M_\odot$ black hole spiral into $10^6$ $M_\odot$ black hole; (iv) galactic compact binaries of white dwarfs, neutron stars, stellar-mass black holes; and (v) primordial gravitational wave sources, strings, boson stars etc.

The general concept of ASTROD (Astrodynamical Space Test of Relativity using Optical Devices) is to have a constellation of drag-free spacecraft navigate through the solar system and range with one another using optical devices to map the solar-system gravitational field, to measure related solar-system parameters, to test relativistic gravity, to observe solar g-mode oscillations, and to detect gravitational waves.

A baseline implementation of ASTROD is proposed in 1993 and is under concept and laboratory studies since then. The mission concept is to have two spacecraft in separate solar orbits, each carrying a payload of a proof mass, two telescopes, two 1–2 W lasers, a clock and a drag-free system, together with a similar spacecraft near Earth at one of the Lagrange points L1/L2.[29,30,58] The three spacecraft range coherently with one another using lasers to map solar-system gravity, to test relativistic gravity, to observe solar g-mode oscillations, and to detect gravitational waves. Distances between spacecraft depend critically on solar-system gravity (including gravity induced by solar oscillations), underlying gravitational theory and incoming gravitational waves. A precise measurement of these distances as a function of time will enable the cause of a variation to be determined. After 2.5 years, the inner spacecraft completes 3 rounds about the Sun, the outer spacecraft 2 rounds, and the L1/L2 spacecraft (Earth) 2.5 rounds. At this stage two spacecraft will be on the other side of the Sun, as viewed from the Earth, in the optimum configuration for the Shapiro time delay experiment.

The shot noise sensitivity limit in the strain for gravitational-wave detection is inversely proportional to $P^{1/2}l$ with $P$ the received power and $l$ the distance. Since $P$ is inversely proportional to $l^2$ and $P^{1/2}l$ is constant, this sensitivity limit is independent of the distance. Conservatively, if LISA accelerometer noise goal is taken for ASTROD, the sensitivity at low frequency is about 30 times lower (the ratio of arm lengths) than LISA.

Since ASTROD is in a time frame later than LISA, if the absolute metrological accelerometer / inertial sensor is developed, there is a potential to improve. ASTROD has a large variation in its triangular formation with arm length varying up to about 2 AU. This demands optical frequency synthesis to match the relative Doppler shift. The weaker light phase locking for amplification is already demonstrated for 2 pW at Tsing Hua University[59,60] and for 40 fW in JPL[61], and is already good for this distance. With the re-focus of deep space exploration, we proposed a dedicated GW mission as follows.[12,14]

ASTROD-GW (ASTROD [Astrodynamical Space Test of Relativity using Optical Devices] optimized for GW detection) is an optimization of ASTROD to focus on the goal of detection of GWs. The detection sensitivity is shifted 52 times toward larger wavelength compared to that of LISA. The scientific aim is focused for gravitational wave detection at low frequency. The mission orbits of the 3 spacecraft forming a nearly equilateral triangular array are chosen to be near the Sun-Earth Lagrange points L3, L4 and L5 (Fig. 1). The 3 spacecraft range interferometrically with one another with arm length about 260 million kilometers. After mission-orbit optimization,[14,62] the changes of arm length are less than 0.0003 AU or, fractionally, less than $\pm 10^{-4}$ in ten years, and the Doppler velocities for the three spacecraft are less than ±4 m/s. These parameters are consistent with those of LISA and a number of technologies developed by LISA could be applied to ASTROD-GW. For the purpose of primordial GW detection, a 6-S/C formation for ASTROD-GW will be used for correlated detection of stochastic GWs.

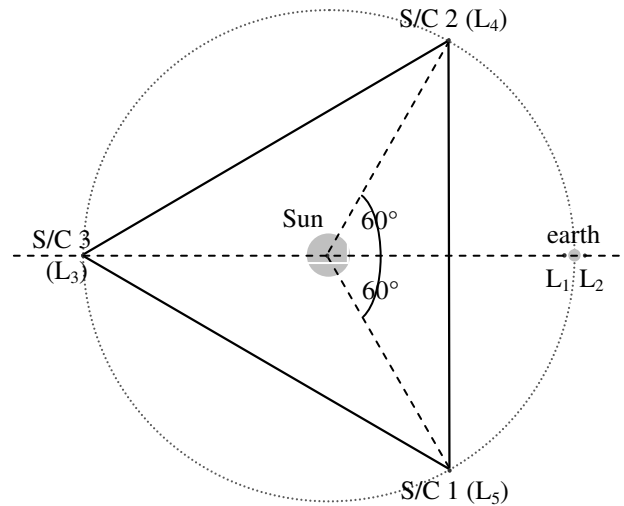

Fig. 1. Schematic of ASTROD-GW mission orbit design.

Since the arm length is longer than LISA by 52 times, with 1-2 W laser power and LISA acceleration noise, the strain sensitivity of ASTROD-GW is 52 times lower than LISA, and is better than LISA and Pulsar Timing Arrays (PTPs) in the frequency band 100 nHz - 1 mHz. ASTROD-GW will complement LISA and PTP in exploring black hole co-evolution with galaxies and GW backgrounds in the important frequency range 100 nHz - 1 mHz.

Recent studies[63,64] of both stochastic GW background and resolvable source from massive black hole binary systems during galaxy co-evolution indicate that they would be detectable with Pulsar Timing Arrays [PTAs] and LISA. With better sensitivity in the frequency bandwidth 100 nHz - 1 mHz, ASTROD-GW is in a unique position to complement and improve on PTAs and LISA for this goal.

For direct detection of primordial (inflationary, relic) GWs in space, one may go to frequencies lower or higher than LISA bandwidth,[65] where there are potentially less foreground astrophysical sources[66] to mask detection. DECIGO[32] and Big Bang Observer[33] look for GWs in the higher frequency range while ASTROD, ASTROD-GW and Super-ASTROD look for GWs in the lower frequency range.

Super-ASTROD (Super Astrodynamical Space Test of Relativity using Optical Devices)[31] is a mission concept with 3-5 spacecraft in 5 AU orbits together with an Earth-Sun L1/L2 spacecraft ranging optically with one another to probe primordial gravitational-waves with frequencies 100 nHz - 1 mHz, to test fundamental laws of spacetime and to map the outer solar system.

DECIGO[32] (DECi-hertz Interferometer Gravitational wave Observatory) is the future Japanese space gravitational wave antenna. It aims at detecting various kinds of gravitational waves between 1 mHz and 100 Hz frequently enough to open a new window of observation for gravitational wave astronomy. The preconceptual design of DECIGO consists of three drag-free satellites, 1000 km apart from each other, whose relative displacements are measured by a Fabry–Perot Michelson interferometer.

Big Bang Observer (BBO) is of LISA type orbit but with 2-5 × $10^4$ km arm length. Both BBO and DECIGO aim at better deci-Hz sensitivity. The Big Bang Observatory is a follow-on mission to LISA, a vision mission of NASA's "Beyond Einstein" theme. BBO will probe the frequency region of 0.01–10 Hz, a region between the measurement bands of the presently funded ground- and space-based detectors. Its primary goal is the study of primordial gravitational waves from the era of the big bang, at a frequency range not limited by the confusion noise from Galactic compact binaries.

In order to separate the inflationary GWs from GWs of the merging binaries, DECIGO and BBO will identify and subtract the signal in its detection band from every merging neutron star and black hole binary in the universe. They will also extend LISA's scientific program of measuring waves from the merging of intermediate-mass black holes at any redshift, and will refine the mapping of space-time around supermassive black holes with inspiraling compact objects. The spectral strain sensitivity of BBO and DECIGO at 0.1 Hz is planned to be ~ $10^{-24}$ $Hz^{-1/2}$.

## 4. Dark energy and GW luminosity distance

Type SN IA supernova observations[68-70] indicate that our universe is currently undergoing an accelerated expansion. This indicates that the cosmological constant is not zero in the framework of general relativity and has stimulated a lot of researches on dark energy. The issue is whether 'cosmological constant' can vary – dark energy issue.[71] In the case of scalar field models, the issue becomes what is the value of $w(\phi)$ in the scalar field equation of state:

$$w(\phi) = p(\phi) / \rho(\phi), \qquad (4)$$

where $p$ is the pressure and $\rho$ the density. For cosmological constant, $w = -1$. From cosmological observations, our universe is close to being flat. In a flat Friedman Lemaître-Robertson-Walker (FLRW) universe, the luminosity distance is given by

$$d_L(z) = (1+z) \int_{0 \to z} (H_0)^{-1} [\Omega_m(1+z')^3 + \Omega_{DE}(1+z')^{3(1+w)}]^{-(1/2)} dz', \quad (5)$$

where $H_0$ is Hubble constant, $\Omega_{DE}$ is the present dark energy density parameter, and the equation of state of the dark energy $w$ is assumed to be constant. In the case of non-constatnt $w$ and non-flat FLRW universe, similar but more complicated expression can be derived. Here we show (5) for illustrative purpose. From the observed relation of luminosity distance vs. redshift $z$, the equation of state $w$ as a function of redshift $z$ can be solved for and compared with various cosmological models. Dark energy cosmological models can be tested this way. Luminosity distance from supernova observations and from gamma ray burst observations vs. redshift observations are the focus for the dark energy probe.

Gravitational waveforms of black hole binaries give precise, gravitationally calibrated distances to high redshift. The inspiral signals of these binaries can serve as standard candles.[72,73] Observations of massive black hole coalescences at cosmological distances by space-based detectors facilitate an accurate determination of the distance to the source. Basically the luminosity distance is given by amplitude, chirp time and orbital frequency

$$\text{Luminosity distance} \approx \text{velocity of light} \times \text{frequency}^{-2} \times t_{chirp}^{-1} \times \text{amplitude}^{-1}. \quad (6)$$

The amplitude, frequency, and chirp rate of the binary can be measured from gravitational wave observation, therefore the luminosity distance can be inferred. The redshift $z$ cannot be inferred from the observed signal for a signal with a redshift z and a chirp mass M looks identical to a signal with no redshift and a chirp mass of M/(1 + z).

For LISA, the accuracy of luminosity distance determination for high redshift event is about 0.1-10 % due to high signal to noise ratio.[74] To use the luminosity distances for determining the equation of state for dark energy (5), one has to obtain redshifts of the host galaxies. High signal to noise ratio gives high angular resolution which facilitates the determination of optical association and redshift. LISA as a dark energy probe will be able to determine the equation-of-state parameter to a couple of a percent.[74]

For ASTROD-GW with 52 times less strain noise, the determination of luminosity distance from high redshift events will be better than 1 %. Luminosity distance together with redshift measurement determines the equation of state of dark energy (6). ASTROD-GW will be able to determine this equation to 1 % or better with the only limitation coming from weak lensing. With more study and observation, the limitation from weak lensing will be clearer and hopefully suppressed to certain extent.

The determination of dark energy equation tests various dark energy models. We illustrate this with two models. In the quintessential inflation, both inflation and dark energy (quintessence) are described by the same scalar field.[75,76] In this scenario, during the radiative scheme, the scalar field equation of state locks to $w(\phi) \approx -1$ and behaves like an effective cosmological constant; during the current epoch of accelerated expansion $w(\phi) \leq -0.9$ and could be 5-10 % different from -1. One percent accuracy in the dark energy equation would be able to distinguish this model from the cosmological constant model. This scenario on dark energy together with other scenarios could be empirically tested in the ASTROD-GW mission. As a second example, to distinguish the Yang-Mills dark energy models,[77] high redshift observations are needed.[78] Currently, only low redshift data have been accumulated with an upper limit up to z ~ 2. With LISA and ASTROD-GW, observations can be extended to z ~ 20 and beyond, this will enable various yang-Mills dark energy models to be distinguished.

**5. Inflation and inflationary GWs**

Inflation is a rapid accelerated expansion which set the initial moments of the Big Bang Cosmology.[79-84] This expansion drives the universe towards a homogeneous and spatially flat geometry that accurately describes the average state of the universe. The quantum fluctuations in this era grow into the galaxies, clusters of galaxies and temperature anisotropies of the cosmic microwave background.[85-90] Although modern inflation is originated from efforts of unification, its mechanism remains unclear. The quantum fluctuations in the spacetime geometry in the inflationary era generate GWs which would have imprinted tensor perturbations on the Microwave Background Radiation anisotropy. The analysis of five-year data of WMAP did not discover these tensor perturbations and showed that, combined with BAO and SN data, the scalar index is $n_s = 0.960 \pm 0.013$ and the tensor-to-scalar perturbation ratio $r$ is less than 0.20 (95% CL).[47] The recent release of the analysis of seven-year data is consistent with the five-year results.[91] More than one order-of magnitude better accuracy in CMB polarization observation is expected from PLANCK mission[48] launched on May 14, 2009. Dedicated CMB polarization observers like B-Pol,[92] EPIC[93] and LiteBIRD[94] missions would improve the sensitivity further by one order-of magnitude. This development would put the measurable $r$ in the range of $10^{-3}$ corresponding to inflation energy scale of about $10^{15}$-$10^{16}$ GeV.

If the tensor mode is detected in the CMB polarization observation, it could leads to two possible causes: (i) due to primordial gravitational waves;[95-96] (ii) due to (pseudo)scalar-photon interaction in the CMB propagation.[97] These possibilities may be distinguishable by detailed models and comparison with better observations.[97] Suppose we have the possibility (i), more on inflationary physics can only be probed by direct gravitational-wave detection since next-generation CMB polarization observation will be limited by cosmic variances and cosmic shears.[92-94]

Smith *et al.*[95-96] connected CMB constraints to the amplitude and tensor spectral index of the inflationary gravitational-wave background (IGWB) at space GW detectors' frequencies for six classes of models of inflation by directly solving the inflationary equations of motion and predicted $\Omega_g$ to be above $10^{-17}$. As the errors on $n_s$ shrink on both sides then, depending on the central value for $n_s$, each of the six models analyzed here may eventually predict minimum amplitude for inflationary GW background. The prediction of the amplitude of inflationary gravitational waves depends on various scenarios in the inflation and varies from authors to authors. One recent estimate[98] ranges up to $10^{-10}$.

The instrument sensitivity goals of DECIGO[32], Big Bang Observer[33-34] and ASTROD-GW[12,14] (correlation detection) all reach $10^{-17}$ in terms of critical density. The main issue is the level of foreground and whether foreground could be separated.

**6. Discussion and Outlook**

We have presented a complete frequency classification of gravitational waves according to their detection methods. Although there is no direct detection of gravitational waves yet, several bands are amenable to direct detection. Direct detection may first come in the high frequency band or in the very low frequency band. The detection in the low frequency band may have the largest signal to noise ratios. This will enable the detailed study of black hole co-evolution with galaxies and the dark energy issue.

Foreground separation and correlation detection method need to be investigated to achieve the sensitivities $10^{-17}$-$10^{-18}$ for $\Omega_{gw}$. With current technology development, we are in a position to explore deeper into the origin of gravitation and our universe. The current and coming generations are holding such promises.

**Acknowledgements**

I am grateful to Swara Ravindranath and Varun Sahni for helpful discussions on co-evolution of massive black holes with galaxies and quintessential inflation. This work is supported in part by the National Science Council (NSC 98-2112-M-007-009).